\documentclass{article}
\usepackage{spconf,amsmath,graphicx}
\usepackage{amsfonts}
\usepackage{url}
\usepackage{hyperref}
\hypersetup{
  colorlinks   = true, 
  urlcolor     = black, 
  linkcolor    = black, 
  citecolor   = black 
}
\usepackage{booktabs}
\usepackage{float}
\usepackage{siunitx}
\title{Neural Percussive Synthesis Parameterised by High-Level Timbral Features}
\name{Ant\'onio Ramires\textsuperscript{1}, Pritish Chandna\textsuperscript{1}, Xavier Favory\textsuperscript{1}, Emilia G\'omez\textsuperscript{1,2}, Xavier Serra\textsuperscript{1}\thanks{This work is partially funded by the European Union’s Horizon 2020 research and innovation programme under the Marie Skłodowska-Curie grant agreement No765068, MIP-Frontiers. This work is partially supported by the Towards Richer Online Music Public-domain Archives (TROMPA) project. The TITANX used for this research was donated by the NVIDIA Corporation. }}
\address{\textsuperscript{1}Music Technology Group, Universitat Pompeu Fabra, Barcelona, Spain\\
\textsuperscript{2}Joint Research Centre, European Commission, Seville, Spain}
\begin{document}
%
\maketitle
\begin{abstract}
We present a deep neural network-based methodology for synthesising percussive sounds with control over high-level timbral characteristics of the sounds. 
This approach allows for intuitive control of a synthesizer, enabling the user to shape sounds without extensive knowledge of signal processing. 
We use a feedforward convolutional neural network-based architecture, which is able to map input parameters to the corresponding waveform. 
We propose two datasets to evaluate our approach on both a restrictive context, and in one covering a broader spectrum of sounds.
The timbral features used as parameters are taken from recent literature in signal processing.
We also use these features for evaluation and validation of the presented model, to ensure that changing the input parameters produces a congruent waveform with the desired characteristics.
Finally, we evaluate the quality of the output sound using a subjective listening test. 
We provide sound examples and the system's source code for reproducibility.
\end{abstract}
%

\section{Introduction}
\label{sec:intro}


Percussion is one of the main components in music and is normally responsible for a song's rhythm section. Classic percussion instruments create sound when struck or scraped; however new electronic instruments were developed for generating these sounds either through playing prerecorded samples or through synthesising them. These are called drum machines and became very popular for electronic music~\cite{tr808music}. However, these early drum machines did not provide much control over the generation of the sounds. With the developments in digital audio technology and computer music, new drum machines were hand-designed using expert knowledge on synthesis techniques and electronic music production.

With the success of deep learning, several innovative generative methodologies have been proposed in the recent years. These methodologies include Generative Adversarial Networks (GANs)~\cite{gans}, Variational Autoencoders (VAEs)~\cite{vae} and autoregressive networks~\cite{oord2016wavenet,engelNsynth}. In the audio domain, such methodologies have been applied for singing voice~\cite{chandna2019wgansing}, instrumental sounds~\cite{engelNsynth} and drum sound generation~\cite{NeuralDrumMachine}. However, in the case of percussive sounds, the proposed methods only allow the user to navigate in non-intuitive high dimensional latent spaces.

The aim of our research is to create a single-event percussive-sound synthesizer that can be intuitively  controlled by users, despite their sound design knowledge. This requires both a back end of a generative model that is able to map the user controls to the output sound and a front end user interface. In this paper, we propose a generative methodology based on the Wave-U-Net architecture~\cite{stoller2018wave}. Our method maps high-level characteristics of sounds to the corresponding waveforms.
The use of these features is aimed at giving the end-user intuitive control over the sound generation process. We also present a dataset of \num{10000} percussive one-shot sounds collected from Freesound~\cite{font2013freesound}, curated specially for this study.

The source code for our model is available online\footnote{\url{https://github.com/pc2752/percussive_synth}}, as are sound examples\footnote{\url{https://pc2752.github.io/percussive_synth/}}, showcasing the robustness of the models.
\section{Generative Models For Audio}
\label{sec:rel_work}

In the audio domain, several generative models have been proposed over the recent years. In the context of music, generative models have shown success specially in creating pitched instrumental sounds, when conditioned on musical notes. A pioneering work on this field was NSynth~\cite{engelNsynth}. This synthesizer is based on the  
wavenet vocoder~\cite{shen2018natural} an autoregressive architecture, which, while capable of generating high quality sounds is very resource intensive. 
Several alternate architectures have been used for the generation of musical notes, based on GANs~\cite{advaudiosynth,engel2018gansynth}, VAEs~\cite{esling2018bridging,NeuralDrumMachine,manytomany}, Adversarial AEs~\cite{bitton} and AEs with Wavenet~\cite{engelNsynth}. 

For percussive sound synthesis, the most relevant work is the Neural Drum Machine~\cite{NeuralDrumMachine}, which uses a Conditional Wasserstein Auto Encoder~\cite{tolstikhin2017wasserstein}, trained on the magnitude component of the spectogram of percussive sounds coupled with a Multi-Head Convolutional Neural Network for reconstructing the audio from the spectral representation. Principal Component Analysis is used on the low-dimensional representation learned by the AE to select the \num{3} most influential dimensions of the \num{64} dimensions of the embedding. These are provided to the user over a control interface. However these parameters controlled by the user are abstract and are not shown to be perceptually relevant or semantically meaningful.

In our case, we wish to directly map a chosen set of features to the output sound. The WaveNet~\cite{oord2016wavenet} architecture has been shown to generate high quality waveforms conditioned on input features. However, the autoregressive nature of the model makes it resource extensive and the short nature of percussive sounds do not require the use of a long temporal model. Therefore, for our study, we decided to use the Wave-U-Net~\cite{stoller2018wave} architecture, which has been shown to effectively model waveforms in the case of source separation and follows a feedforward convolutional architecture, making it resource efficient. The model takes as input the waveform of the mixture of sources, downsamples it through a series of convolutional operations to generate a low dimensional representation and then upsamples it through linear interpolation followed by convolution to the output dimensions. There are concatinative connections between the corresponding layers of the upsampling and downsampling blocks. In our work, we adapt this architecture to fit the desired use case.





\section{Timbral features}
\label{sec:sem_feat}
For our end goal, we require semantically meaningful features that can allow for intuitive control of the synthesizer. In the field of Music Information Retrieval, a strong effort has been put on developing hand crafted features which can characterise sounds. These features enable users to retrieve sounds or music from large audio collections by automatically describing them according to their timbre, their mood, or other characteristics which are easy to understand by users. For our purpose, we need features pertaining to timbre. We understand timbre as pertaining to perceptual characteristics of sounds analogous to colour or quality. Control over such features would enable the user to intuitively shape sounds.  

A set of such features have been proposed in~\cite{pearce2017timbral}, where recurrent query terms, related to timbral characteristics, used for searching sounds in large audio databases were identified. Regression models were developed by mapping user-collected ratings to timbral characteristics, which quantify semantic attributes. These are hardness, depth, brightness, roughness, boominess, warmth and sharpness.
The work proposes feature extractors pertaining to these query terms and we use an open-source implementation of the same\footnote{\url{https://github.com/AudioCommons/ac-audio-extractor}}. For the rest of this paper, we refer to the \num{7} features extracted by these extractors as timbral features.


Another relevant characteristic which is commonly present in drum synthesizers and music makers are used to work with is the temporal envelope of the sound. This feature describes the energy of the sounds over time and is normally available to users in drum synthesizers as a set of \textit{attack} and \textit{decay} controls. We use an open-source implementation of the envelope algorithm described in \cite{zolzer2008digital}, present in the Essentia library~\cite{essentia}. An attack time of \SI{5}{\milli\second} and a release time of \SI{50}{\milli\second} was used to generate a smooth curve which matched the sound energy over time.
It must be noted that the timbral features described previously are summary features, i.e. have a single value for each sound while the envelope is time evolving and of the same dimensions as the waveform.

\section{DATASET CURATION}
\label{sec:dataset}
We curated two datasets in order to train our models in different scenarios.
The first consists of sounds taken from Freesound, a website which hosts a collaborative collection of Creative Commons licensed sounds\footnote{\url{https://freesound.org}}~\cite{font2013freesound}.
We performed queries to the database with the name of percussion instruments as keywords in order to retrieve a set of percussive sounds, with a limit on effective duration of \SI{1}{\second}.
We then conducted a manual verification of these sounds\footnote{We developed an annotation tool available at this repository \url{https://github.com/xavierfav/percussive-annotator}.}:  to select the ones that were containing one single event, and were of appreciable quality in the context of traditional electronic music production. 
This process created a dataset of around \num{10000} sounds, containing instruments such as kicks, snares, cymbals and bells. The dataset is publicly available in a Zenodo repository\footnote{\url{https://zenodo.org/record/3665275}}. For the rest of this paper, we refer to this dataset as FREESOUND.

A second dataset was created by aggregating about \num{5000} kick drum one-shot samples from our personal collections, originating mostly from commercial libraries.
This type of sounds are often of high quality, annotated and contain only one event which makes it very handy to construct a dataset of isolated sounds, suiting our needs for training our model in a restricted context. We refer to this dataset as KICKS.

The aim of creating two datasets was to understand if our method could be applicable for synthesising a wide variety of percussion sounds, or if it was more appropriate to focus on synthesising only one type of sounds, in this case the kick drum.



\section{Methodology}
\label{sec:system}

\begin{figure}[ht]
\centering

\includegraphics[width=0.5\textwidth]{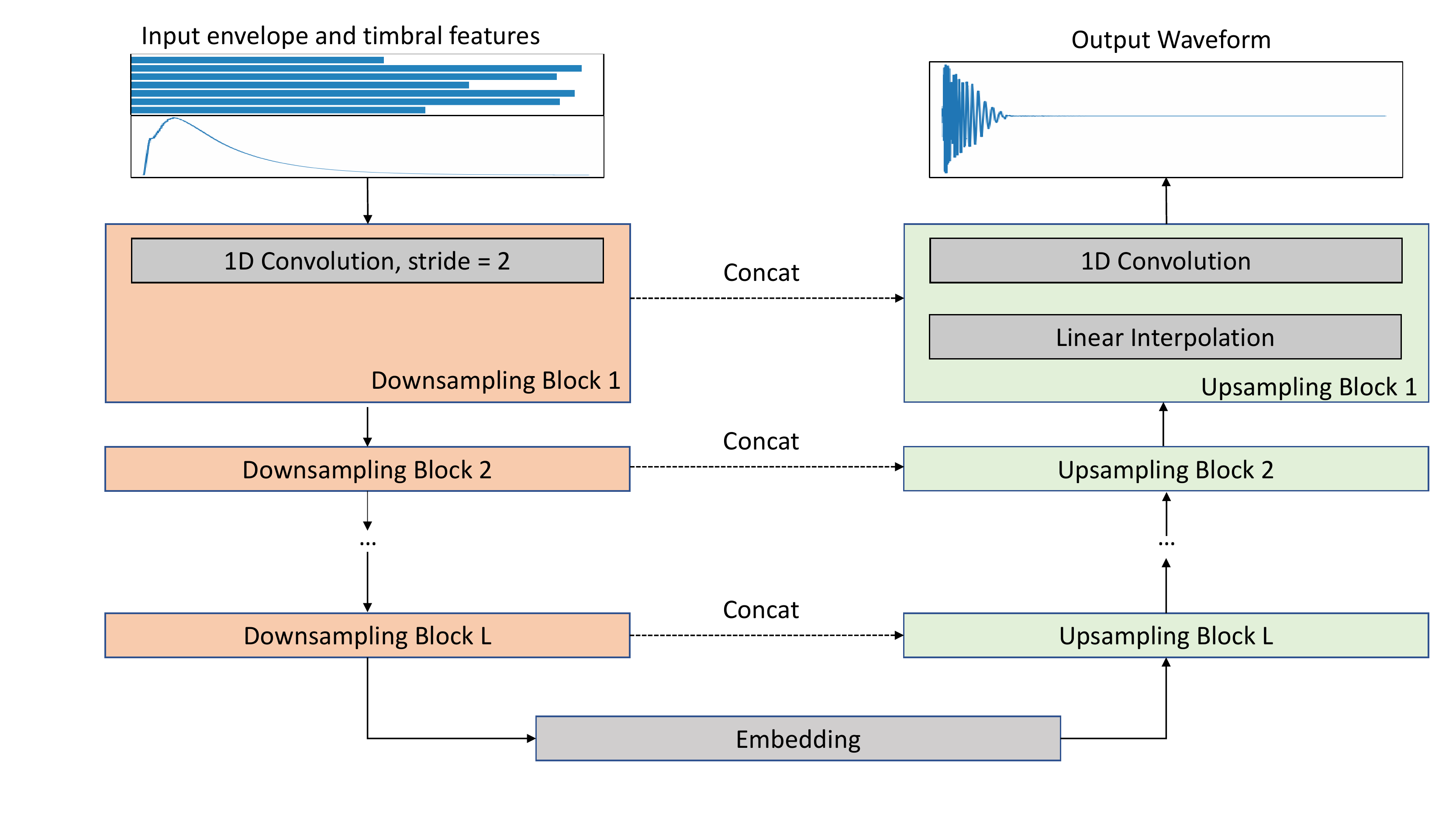}
 \caption{The proposed architecture, with $K=15$ layers.}
 \label{fig:archi}
\end{figure}

We aim to model the probability distribution of the waveform ${x}$ as a function of the timbral features $fs$ and the time-domain envelope $e$. To this end we use a feedforward convolutional neural network as a function approximator to model $P(x|fs,e)$. We use a U-Net architecture, similar to the one used by \cite{stoller2018wave}, which has been shown to effectively model the waveform of an audio signal. Our network takes the envelope as input and concatenates to it the timbral features, $fs$, broadcast to the input dimensions, as done by \cite{oord2016wavenet}. As shown in Figure \ref{fig:archi}, downsampling is done via a series of convolutions with stride \num{2}, to produce a low-dimension embedding. We use a filter length of \num{5} and double the number of filters after each \num{3} layers, starting with \num{32} filters. A total of \num{15} layers are used in the encoder, leading to an embedding of size \num{512}.
We upsample this low dimensional embedding sequentially to the output $\hat{x}$, using linear interpolation followed by convolution. This mirrors the approach used by \cite{ chandna2019wgansing,stoller2018wave} and has been shown to avoid high frequency artefacts which appear while upsampling with transposed convolutions. As with the U-Net, there are connections between the corresponding layers of the encoder and decoder, as shown in Figure \ref{fig:archi}.

We initially used a simple reconstruction loss function, shown in equation \ref{eq:loss1} to optimise the network.

\begin{equation}
\mathcal{L}_{recon} = \mathbb{E} [\|\hat{x} - x\|_1] 
\label{eq:loss1}
\end{equation}

While this resulted in a decent output, we noticed that the network was able to reproduce the low frequency components of the desired sound, but lacked details in high frequency components. To rectify this, we added a short time fourier transform (\textit{STFT}) based loss, similar to \cite{sahai2019spectrogram}. 
This loss is shown in equation \ref{eq:loss2}. 

\begin{equation}
\mathcal{L}_{stft} = \mathbb{E} [\|STFT(\hat{x}) - STFT(x)\|_1] 
\label{eq:loss2}
\end{equation}

The final loss of this network is shown in equation \ref{eq:loss3}. 
\begin{equation}
\mathcal{L}_{final} = \mathcal{L}_{recon} + \lambda\mathcal{L}_{stft}
\label{eq:loss3}
\end{equation}

Where $\lambda$ is the weight given to the high frequency component of the reconstruction. 





\section{Experiments}
\label{sec:results}
\subsection{Data Pre-processing}
\label{sec:feats}
All sound were downsampled to a sampling rate of \SI{16}{\kilo\hertz} and silences were removed from the beginning and end of the sounds. Following this, we calculated the timbral features and envelope described in section \ref{sec:sem_feat} and then zero-padded at the end of the sound to \num{16000} samples. The features were normalised using min-max normalisation, to ensure that the inputs were within the range \SIrange{0}{1}.  

\subsection{Training the network}
The network was trained using the Adam optimiser \cite{kingma2014adam} for \num{2500} epochs with a batch size of \num{16}. We use \SI{90}{\percent} of the data for training and \SI{10}{\percent} for evaluation.
The \textit{STFT} used for the $\mathcal{L}_{stft}$ loss function is calculated over \num{1024} samples and a hopsize of \num{512}. With the given sampling rate, this led to a frequency resolution of \SI{16.125}{\hertz} per bin. We evaluate the model with three losses: the $\mathcal{L}_{recon}$ loss, henceforth referred to as WAVE; the $\mathcal{L}_{final}$, referred to as FULL; and a version with only the high frequency components of the \textit{STFT} for the $\mathcal{L}_{stft}$, referred as HIGH. This last model uses \textit{STFT} components above \SI{650}{\hertz} or \num{40} bins as traditional kick synthesizers model a kick sound via a low frequency sinusoid, generally below \SI{650}{\hertz} with some high frequency noise. We use $\lambda=0.5$ for our experiments.

\subsection{Evaluation}
\begin{figure}
\centering

\includegraphics[width=0.5\textwidth]{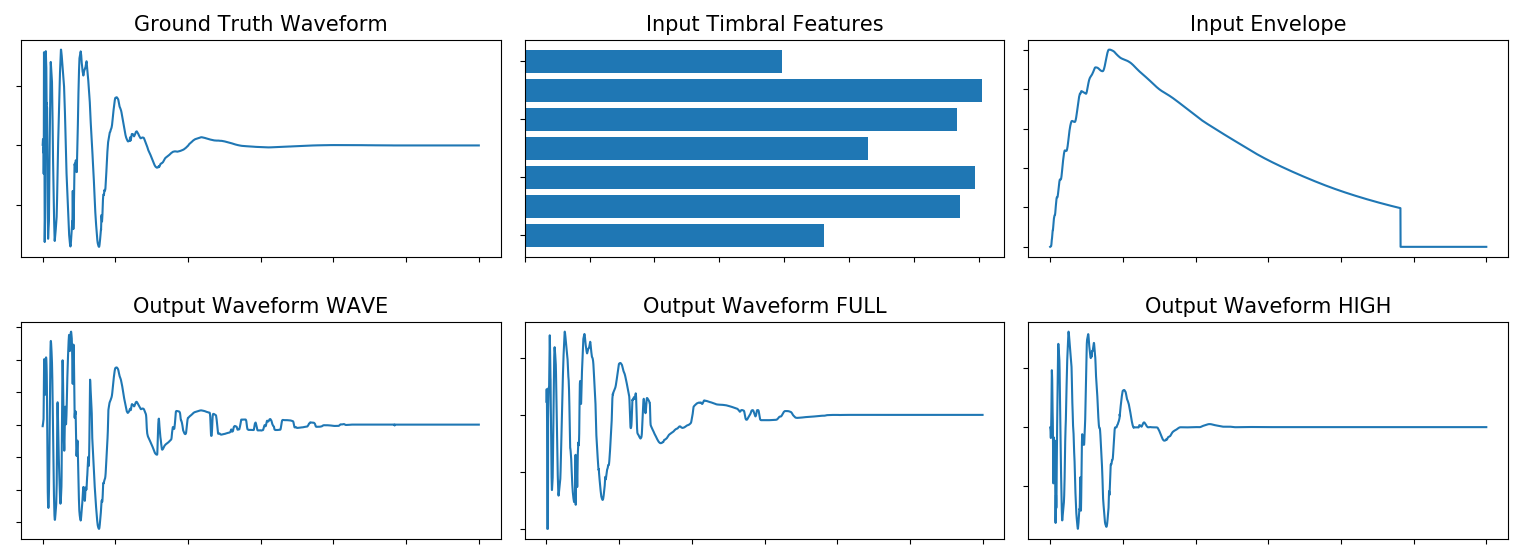}
 \caption{A sample of the input envelope and features and the output waveforms for the various models for the KICK dataset}
 \label{fig:outputs_kicks}

\end{figure}  
The proposed models need to be evaluated in terms of the perceived audio quality and the coherence of timbral features between the input and the output. 
A preliminary assessment of the quality of reconstruction can be made by looking at the output waveforms, shown in Figure \ref{fig:outputs_kicks} for a sample from the test set of the KICKS dataset. Although the reconstruction seems to be visually accurate for the three models, the perceived quality of the audio is a subjective metric that cannot be judged by simply looking at the plots. 
We can objectively assess the coherence of the timbral features used as input to the model. More importantly, we want to assess that a change in these features leads to a corresponding change in the output. 

To this end, we vary each individual timbral feature while maintaining the other features constant. We then check the accuracy of the output waveform via the same feature extractors used for training. For each individual feature, we set values of \textit{low}, corresponding to \num{0.2} over the normalised scale, \textit{mid}, corresponding to \num{0.5} and \textit{high}, corresponding to \num{0.8}. The respective outputs for such models are termed $\hat{x}_{low}^i$, $\hat{x}_{mid}^i$ and $\hat{x}_{high}^i$ and their corresponding features are $fs_{low}^i$, $fs_{mid}^i$ and $fs_{high}^i$ for the $i^{th}$ feature. 
For coherent modelling, the models should follow the order
$fs_{high}^i >fs_{mid}^i>fs_{low}^i$. We asses the accuracy of this order in three tests, $E1$, which checks the condition $fs_{high}^i>fs_{low}^i$, $E2$, which checks $fs_{high}^i>fs_{mid}^i$ and $E3$, which checks $fs_{mid}^i>fs_{low}^i$ over all values of $i$. The accuracy of the models over these tests is shown in Table \ref{table:models} and a feature wise summary is shown in Table \ref{table:feats}. 

\begin{table}[h]
\centering

\begin{tabular}{l c c c c }

 &  &  & Accuracy &  \\ 
Dataset & Model & E1 & E2 & E3 \\ \hline
 & WAVE & 0.601 & 0.569 & 0.552 \\
FREESOUND & HIGH & 0.649 & 0.601 & 0.657 \\
 & \textbf{FULL} & \textbf{0.825} & \textbf{0.758} & \textbf{0.780} \\ \hline
 & WAVE & 0.805 & 0.722 & 0.722 \\ 
KICKS & HIGH & 0.876 & 0.789 & 0.769 \\ 
 & \textbf{FULL} & \textbf{0.920} & \textbf{0.814} & \textbf{0.798} \\ 
\end{tabular}
\caption{Objective verification of feature coherence across models and datasets.}
\label{table:models}
\end{table}

It can be seen that the FULL model, followed by the HIGH, are the most efficient at mapping the input features to the output waveform in terms of feature coherence, but all three models do maintain this coherence to a high degree. 

\begin{table}[h]
\centering

\begin{tabular}{l | c c c | c  c c }
&  \multicolumn{3}{c|}{FREESOUND}  & \multicolumn{3}{c}{KICKS} \\
Feature & E1 & E2 & E3 & E1 & E2 & E3 \\ 
\hline
Boominess   & 0.98 & 0.82 & 0.98 & 0.96 & 0.86 & 0.95  \\ 
Brightness  & 0.99 & 0.99 & 1.00 & 0.99 & 0.98 & 0.84  \\ 
Depth       & 0.94 & 0.65 & 0.94 & 0.99 & 0.89 & 0.94  \\ 
Hardness    & 0.64 & 0.66 & 0.59 & 0.85 & 0.61 & 0.79   \\ 
Roughness   & 0.63 & 0.59 & 0.57 & 0.84 & 0.80 & 0.62  \\ 
Sharpness   & 0.63 & 0.77 & 0.45 & 0.90 & 0.91 & 0.54  \\ 
Warmth      & 0.92 & 0.79 & 0.91 & 0.88 & 0.61 & 0.87   \\ 

\end{tabular}
\caption{Objective verification of the accuracy on feature coherence for the best performing models for each dataset.}
\label{table:feats}
\end{table}
While feature coherence is maintained for features like boominess, brightness, depth and warmth for the full dataset, the models are less consistent in terms of hardness, roughness and sharpness, particularly true for the FREESOUND dataset.

\begin{figure}[H]
\centering
\includegraphics[width=0.48\textwidth]{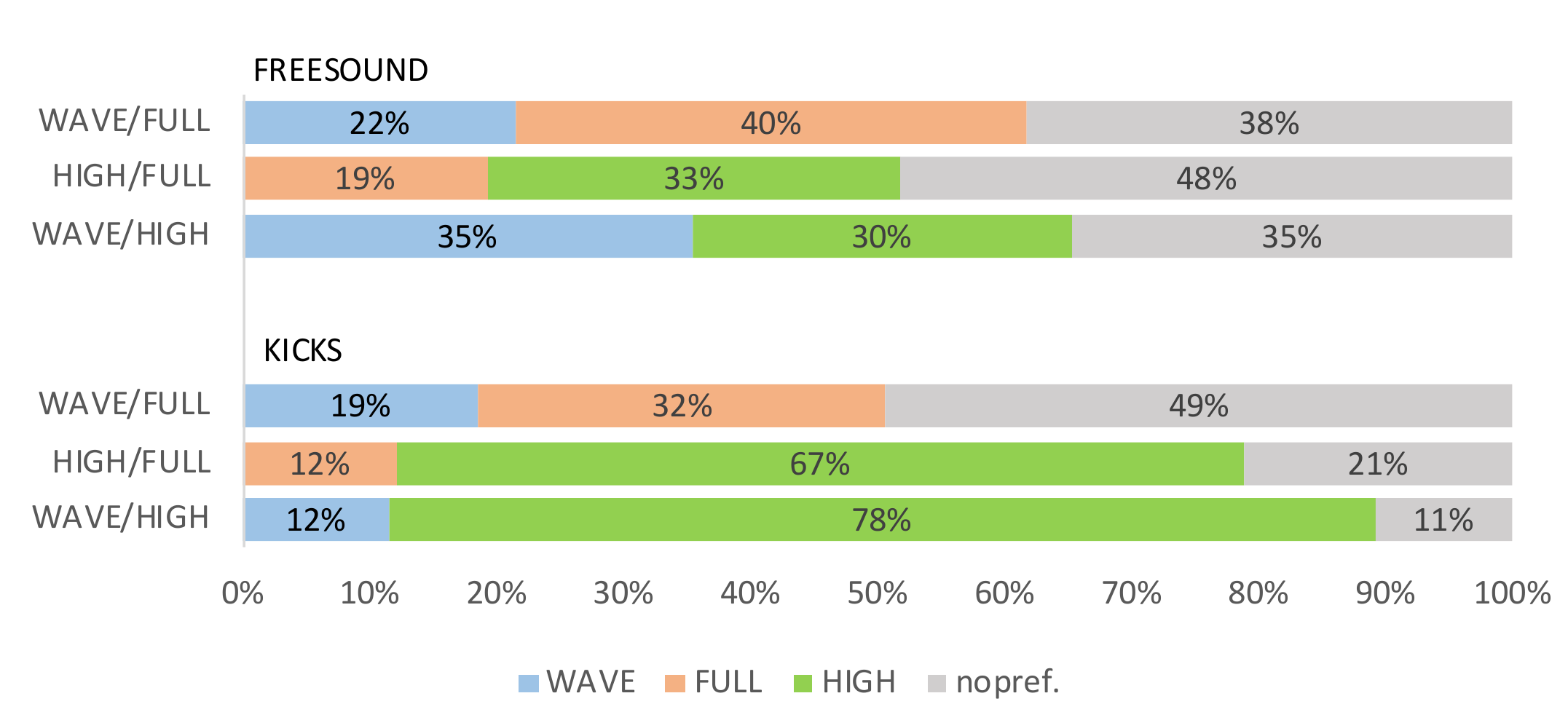}
\vspace{-1em}
 \caption{Results of the listening test, displaying the user preference between loss functions for each of the datasets.}
 \label{fig:surv_res}
\end{figure}

Given the absence of a suitable baseline system, we decided to use an online AB listening test that compared the models amongst themselves and a reference for subjective evaluation of quality. The participants of the test were presented with \num{15} examples each from both datasets. Each example had two options, A and B from two of the models used for the dataset, along with a reference ground truth audio. There were \num{5} examples each from each of the \num{3} pairs. The participant was asked to choose the audio clip which was closest in quality to the reference audio. There were \num{35} participants in the listening test, the results of which are shown in Figure \ref{fig:surv_res}. 

A clear preference for the HIGH model can be seen, especially for the KICKS dataset. This can be attributed partly to the choice of cutoff frequency used in the model and partly to the diversity of sounds in the FREESOUND dataset. We note the difficulty in assessing audio quality over printed text and encourage the user to visit our demo page and listen to the audio samples for assessment.




\section{Conclusions And Future Work}
\label{sec:conclusion}

In this work, we proposed a method using a feedforward convolutional neural network based on the Wave-U-Net~\cite{stoller2018wave} for synthesising percussive sounds conditioned on semantically meaningful features.

Our final aim is to create a system that can be controlled using high-level parameters, being semantically meaningful characteristics that correspond to concepts casual music makers are familiar with. To this end, we use hand crafted features designed by MIR experts and curate and present a dataset for the purpose of modelling percussive sounds. Via objective evaluation, we were able to verify that the control features do indeed modify the output waveform as desired and quality assessment was done via an online listening test. 

Future work will focus on developing an interface for interacting with the synthesizer, which will allow to evaluate the approach in its context of use, with real users.

\vfill\pagebreak

\bibliographystyle{IEEEbib}
\bibliography{refs}

\end{document}